\documentclass[10pt,journal,compsoc]{IEEEtran}

\ifCLASSOPTIONcompsoc
  \usepackage[nocompress]{cite}
\else
  \usepackage{cite}
\fi

\ifCLASSINFOpdf
    \usepackage[pdftex]{graphicx}
\else
    \usepackage[dvips]{graphicx}
\fi
\usepackage{amsmath}
\usepackage{amssymb}

\newtheorem{definition}{Definition}
\newtheorem{condition}{Condition}

\newcommand{\thirdtextwidth}{\dimexpr0.31\textwidth\relax}

\ifCLASSOPTIONcompsoc
  \usepackage[caption=false,font=footnotesize,labelfont=sf,textfont=sf]{subfig}
\else
  \usepackage[caption=false,font=footnotesize]{subfig}
\fi
\begin{document}
%
\title{UPR: Deadlock-Free Dynamic Network Reconfiguration by Exploiting
Channel Dependency Graph Compatibility}
%
%
%
%

\author{Juan-Jos\'{e}~Crespo,
        Jos\'{e}~L.~S\'{a}nchez,
        Francisco~J.~Alfaro-Cort\'{e}s,
        Jos\'{e}~Flich,
        Jos\'{e}~Duato
\IEEEcompsocitemizethanks{
    \IEEEcompsocthanksitem{
        Juan-Jos\'{e}~Crespo,
        Jos\'{e}~L.~S\'{a}nchez and
        Francisco~J.~Alfaro-Cort\'{e}s are with the Departamento de
        Sistemas Inform\'{a}ticos, Universidad de Castilla-La Mancha,
        Albacete, Spain.

        Email: \{juanjose.gcrespo, jose.sgarcia, fco.alfaro\}@uclm.es
}

    \IEEEcompsocthanksitem{
        Jos\'{e}~Flich and Jos\'{e}~Duato are with the Departament
        d'Inform\`{a}tica de Sistemes i Computadors, Universitat
        Polit\`{e}cnica de Val\`{e}ncia, Valencia, Spain.

        Email: \{jflich, jduato\}@disca.upv.es
}
}}

\IEEEtitleabstractindextext{%
\begin{abstract}
Deadlock-free dynamic network reconfiguration process is usually studied
from the routing algorithm restrictions and resource reservation
perspective.
%
The dynamic nature yielded by the transition process from one routing
function to another is often managed by restricting resource usage in
a static predefined manner, which often limits the supported routing
algorithms and/or inactive link patterns, or either requires additional
resources such as virtual channels.
%
Exploiting compatibility between routing functions by exploring their
associated Channel Dependency Graphs (CDG) can take a great benefit from
the dynamic nature of the reconfiguration process.
%
In this paper, we propose a new dynamic reconfiguration process called
Upstream Progressive Reconfiguration (UPR). Our algorithm progressively
performs dependency addition/removal in a per channel basis relying on
the information provided by the CDG while the reconfiguration process
takes place. This gives us the opportunity to foresee compatible
scenarios where both routing functions coexist, reducing the amount of
resource drainage as well as packet injection halting.
\end{abstract}

\begin{IEEEkeywords}
Interconnection networks, dynamic network reconfiguration, channel
dependency graph, deadlock-freedom.
\end{IEEEkeywords}}

\maketitle

\IEEEdisplaynontitleabstractindextext

\ifCLASSOPTIONpeerreview
\begin{center} \bfseries © 2020 IEEE. Personal use of this material is
permitted. Permission from IEEE must be obtained for all other uses,
in any current or future media, including reprinting/republishing
this material for advertising or promotional purposes, creating new
collective works, for resale or redistribution to servers or lists,
or reuse of any copyrighted component of this work in other works.
\end{center}
\fi
%
\IEEEpeerreviewmaketitle

\IEEEraisesectionheading{\section{Introduction}\label{sec:introduction}}

\IEEEPARstart{H}{igh} performance communication subsystems (a.k.a.
networks) have played a key role during the last two decades. The
race towards exascale systems requires the use of a large amount of
interconnected computing nodes to be able to perform at a rate of
$10^{18}$ operations per second\cite{bergman2008exascale}. Additionally,
new parallel applications such as intelligence processing and analysis
require of high capability supercomputers and large datacenters. As a
consequence, high performance interconnection networks must be designed
and scaled accordingly.

Besides performance, network energy efficiency has become an
important concern. Networks present in exascale systems can
consume $10-20\%$ of the total system power. Future networks are
estimated to consume around $30\%$ of the total supercomputer power
budget\cite{dickov2015self}. Despite the improvements in transmission
technology and the use of different signaling techniques, such complex
interconnection deployments need to be highly robust and flexible,
reflecting the users' needs while reducing performance degradation and
energy consumption.

From the performance point of view, adaptive routing algorithms may
provide the required flexibility for an efficient use of network
resources. On the other hand, the energy efficiency issue is addressed
by existing proposals through \emph{power-aware} network techniques,
which often rely on dynamic link width/frequency, and on/off
links\cite{jin2017survey}. This usually presents a \emph{dichotomy} over
these two aspects. Adaptiveness of routing algorithms can be exploited
to maximize network performance for a particular network topology and/or
application requirements. Besides, it can also be exploited to keep
connectivity between source-destination pairs while switching on/off
links to improve energy efficiency. A balance between these two aspects
will depend on network topology state and users' needs.

Network reconfiguration decouples these two aspects. From one point
of view, routing algorithm flexibility can be fully exploited to
improve network performance. From the other, accounting for disabled
links due to \emph{power-aware} network techniques to improve energy
efficiency is handled by the reconfiguration process itself while
moving from the previous routing function to the desired one. Hence,
network reconfiguration allows the network to be able to adapt itself
dynamically, reflecting changes affecting its state. During the
reconfiguration process, network performance should not experience a
degradation that either prevents or interferes to a great extent with
users applications.

Challenges arise for the reconfiguration process to be carried
out successfully regarding deadlock scenarios. Network deadlock
principles have been studied to a great extent by previous
theories\cite{duato1995necessary,dally1987deadlock}. Briefly, a
deadlock state is reached if packets cannot be routed due to a circular
hold-and-wait dependency on network resources (such as storage
buffers). Reaching a deadlocked state has severe consequences regarding
network availability and performance, which may render the system
unusable unless some action is taken.

Due to the dynamic nature of the reconfiguration process, residual
dependencies activated by packets routed under a previous routing
function may interact with dependencies from packets following a
different routing function imposed by the reconfiguration process. This
interaction can lead to reconfiguration-induced deadlock as long as
packets under the influence of multiple routing functions remain
undelivered, even if the routing functions themselves are deadlock-free.
Theoretical support for reconfiguration-induced deadlock, characterizing
this scenario, was provided by Duato's theory\cite{duato2005theory}.

Deadlock-free network reconfiguration techniques can be classified
in two broad categories. In \emph{static reconfiguration}, packet
injection into the network is halted while network resources
are drained of packets routed according to the old routing
function. Then, packet injection is resumed under the new routing
function\cite{teodosiu1997hardware, schroeder1991autonet,
rodeheffer1991automatic}. These techniques, however, suffer from high
packet latencies and significant packet dropping specially when link
deactivation is performed.

On the other hand, \emph{dynamic reconfiguration} schemes allow
transmission of packets following both the old and the new
routing function.  While reducing the amount of dropped packets
and packet latency, these techniques must be devised to avoid
reconfiguration-induced deadlock discussed earlier, which usually
leads to more sophisticated algorithms\cite{casado2001protocol,
avresky2001dynamic, pinkston2003deadlock, lysne2000fast,
lysne2008efficient, balboni2013optimizing}.

In this paper we present a new dynamic reconfiguration scheme based on
the application of Lysne's methodology for developing deadlock-free
dynamic network reconfiguration processes\cite{lysne2005methodology}. We
call this new scheme \emph{Upstream Progressive Reconfiguration}
(UPR). Our scheme is able to perform the reconfiguration process in a
topology-agnostic manner. Moreover, it does not require additional
resources such as virtual channels or separate escape paths. Among
its features, UPR reduces the amount of channels which have to be
drained of packets and it performs selective injection halting between
communication node pairs if necessary. Thus, injection could be halted
for a period as short as possible for a subset of source-destination
pairs in order to prevent reconfiguration-induced deadlocks.

According to the obtained results, UPR effectively exploits
compatibility between the considered routing functions reducing the
amount of channels requiring packet drainage down to only $14\%$
of the total amount of channels available in the network in some
scenarios. Depending on the compatibility yielded between the old and
the new routing functions, packet injection \emph{selective halting} can
be completely avoided. In other words, communication between all pairs
of processing nodes is allowed during the reconfiguration
process. Results unveil the potential of UPR to be used with adaptive
routing algorithms and \emph{power-aware} techniques.

The rest of this paper is organized as follows. Section
\ref{sec:background} provides the base theoretical concepts and notation
formalisms consistent with Duato's theory\cite{duato2005theory}
and Lysne's methodology\cite{lysne2005methodology}. Section
\ref{sec:related-work} gives a further description of existing dynamic
reconfiguration schemes. In Section \ref{sec:upr} we describe the
UPR scheme proposed. Following, Sections \ref{sec:evaluation} and
\ref{sec:conclusion} provide an exploratory analysis applying our scheme
along some concluding remarks. Finally, Section \ref{sec:future-work}
lays out possible improvements to our proposal as well as ongoing
research and future work.

\section{Background}\label{sec:background}

The reconfiguration process presented in this article
relies on Duato's Theory for Deadlock-free Dynamic Network
Reconfiguration\cite{duato2005theory} and follows Lysne's Methodology
for Developing Deadlock-Free Dynamic Network Reconfiguration
Processes\cite{lysne2005methodology}.

One of the main properties of a dynamic reconfiguration process is
that reconfiguration usually follows a partial order among different
switching elements (a.k.a. routers). In other words, due to the
asynchronous operation of routers, it is not feasible to update the
routing information of all at once, specially for large systems
requiring lots of these elements in a distributed environment.

This behavior is represented as a series of update steps performed
locally at each router by following a partial order with respect to
other steps being finished at different routers. In order to lay this
clearly we have included some well-known definitions which will be used
later.

\begin{definition} An \emph{interconnection network} $I$ is a strongly
    connected directed multigraph, denoted by $I = G(N, C)$, where $N$
    is the vertex set representing the processing element nodes $P$
    and the set of router nodes $RT$ such that $N = P \cup RT$. $C$
    is the arc set representing the channels (and virtual channels)
    which is composed of three disjoint non-empty subsets $C_N = \{c_i :
    Src_C(c_i), Dst_C(c_i) \in RT\}$ , $C_I = \{c_i : Src_C(c_i) \in P$
    and $Dst_C(c_i) \in RT\}$, and $C_D = \{c_i : Src_C(c_i) \in RT$ and
    $Dst_C(c_i) \in P\}$ such that $C = C_I \cup C_N \cup C_D$, $C_{IN}
    = C_I \cup C_N$, and $C_{ND} = C_N \cup C_D$. Here, $Src_C(c_i)$ and
    $Dst_C(c_i)$ represent the source and destination node of channel
    $c_i$ respectively.
\label{def:in}
\end{definition}

\begin{definition} A \emph{routing function} $R: C_{IN} \times P
    \rightarrow \mathcal{P}(C_{ND})$ which returns the alternative
    output channels to send a packet whose head flit is at the head
    of current channel $c_c \in C_{IN}$ willing to reach node $n_d
    \in P$. Therefore, $R(c_c, n_d) = \{c_1, c_2,...,c_m\}$ where
    $c_1,...,c_m \in C_{ND}$ are the candidate channels for the packet's
    head flit to be forwarded.
\label{def:rf}
\end{definition}

\begin{definition}
    For any given routing function $R$ and channel $c \in C$, the
    \emph{local routing function} at channel $c$ is denoted as
    $Local_{R\{c\}}(d) : P \rightarrow \mathcal{P}(C)$ such that
    $\forall c \in C$, $\forall d \in P$, and $\forall c' \in C$, $c'
    \in Local_{R\{c\}}(d)$ if an only if $c' \in R(c,d)$.
\label{def:lrf}
\end{definition}

\begin{definition}
    A \emph{reconfiguration process} $RP$ is defined as the progressive
    upgrade of an initial routing function $R_0$ to another one $R_k$
    in $k$ partially synchronized steps across channels $(k \geq
    1)$. Each of these steps (performed locally at each router) is made
    of a condition and an operation: $step_i = <cond_i, oper_i>$. The
    condition, $cond_i$, is a predicate on the state of the local
    configuration at the router, while the operation, $oper_i$,
    indicates what change to make at $step_i$ to the local routing function
    at channel $c$ so as to migrate from $Local_{R_{i-1}\{c\}}$ to
    $Local_{R_{i}\{c\}}$. A partial order $order_i : \prec_i$ is defined
    on channels to synchronize steps across them. Thus, to carry out
    $step_i$ at any given channel, it has to wait until all channels
    preceding it have started (possibly finished) with $oper_i$. If
    $order_i$ is satisfied, $cond_i$ is checked. In addition, if
    $cond_i$ is satisfied, it starts updating its local routing function
    and informs channels following it in $order_i$ that it has started
    updating.
\label{def:rp}
\end{definition}

\begin{definition} The \emph{channel dependency graph} $CDG$ of a
    routing function $R$ defined on interconnection network $I$ is a
    directed graph $CDG = (C, A)$, where the vertex set $C$ corresponds
    to the set of channels in $I$, and the arc set $A$ consists of pairs
    of channels $(c_i, c_j), c_i, c_j \in C$ such that, according to
    $R$, there is a direct dependency from $c_i$ to $c_j$.
\end{definition}

\begin{definition} The \emph{target channel dependency graph} $TCDG$
    of a routing function $R$ is a directed multi-graph $TCDG_R = (C,
    D)$, where the vertex set $C$ corresponds to the set of channels in
    $I$, and the arc set $D$ consists of tuples $(c_i, c_j, t) : c_i,
    c_j \in C, t \in P$, such that, according to $R$, there is a direct
    dependency from channel $c_i$ to $c_j$ for a given target node $t$.
\label{def:tcdg}
\end{definition}

\begin{definition} The set of \emph{outgoing dependencies} $D^+: C
    \rightarrow D$ for a given channel $c_i$ in $TCDG$ is the set of
    dependencies with source channel $c_i$. Therefore, $D^{+}(c_i) =
    \{(c_i, c_j, t) \in D, \forall c_j \in C$ and $\forall t \in P\}$.
\label{def:d+}
\end{definition}

\begin{definition} Let $T^{+}(c_i)$ be the set of \emph{outgoing
    targets} provided by dependencies in $D^+(c_i)$ for channel $c_i$
    in $TCDG$. Hence, $T^{+}(c_i) = \{t : \forall (c_i, c_j, t) \in
    D^{+}(c_i)\}$.
\label{def:t+}
\end{definition}

\begin{definition} The set of \emph{incoming dependencies} $D^-: C
    \rightarrow D$ for a given channel $c_i$ in $TCDG$ is the set of
    dependencies with destination channel $c_i$. Therefore, $D^{-}(c_i) =
    \{(c_j, c_i, t) \in D, \forall c_j \in C$ and $\forall t \in P\}$
\label{def:d-}
\end{definition}

\begin{definition} Let $T^{-}(c_i)$ be the set of \emph{incoming
    targets} provided by dependencies in $D^-(c_i)$ for channel $c_i$
    in $TCDG$. Hence, $T^{-}(c_i) = \{t : \forall (c_j, c_i, t) \in
    D^{-}(c_i)\}$.
\label{def:t-}
\end{definition}

For the following definitions, we will make use of the local routing
functions $Local_{R\{c\}}$ from Definition \ref{def:lrf}. Notice
that the prevailing routing function across all channels $R_P$ is
unique. Besides $R_P$ is given by the composition ($\odot$) of all
local routing functions at each channel at a given moment (i.e. $R_P =
\odot_{c \in C}Local_{R_{Step_{C}(c)}\{c\}}$) as stated in Definition
12 from Duato's theory\cite{duato2005theory}.

\begin{definition}
    A \emph{local routing function} $Local_{R_b\{c\}}$ is said to
    be target conforming with respect to $R_a$ for a given channel
    $c$, if all of the incoming targets supplied by $R_a$ towards
    channel $c$ (in $TCDG_{R_a}$) are also supplied (outgoing targets)
    by $Local_{R_b\{c\}}$ (in $TCDG_{R_b}$) for channel $c$. Using
    Definition \ref{def:t+}:

    \begin{equation}
        T^{-}_{TCDG_{R_a}}(c) \subseteq T^{+}_{TCDG_{R_b}}(c)
        \label{eq:target-conforming-lrf}
    \end{equation}

\label{def:target-conforming-lrf}
\end{definition}

\begin{figure}[!t]
\centering
\includegraphics[width=\columnwidth]{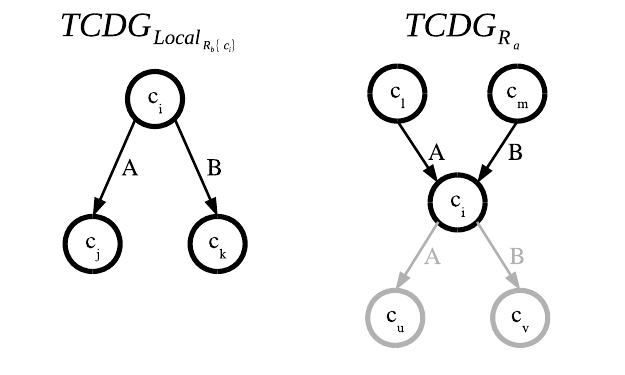}
\caption{Target conforming local routing function $Local_{R_a\{c_i\}}$
for channel $c_i$ with respect to $R_a$. Notice that only the channels
involved within their associated $TCDG$ representation are shown.}
\label{fig:target-conforming-lrf}
\end{figure}

Figure \ref{fig:target-conforming-lrf} shows an example of a target
conforming local routing function. We can see that the local routing
function is a target conforming local routing function due to the
existence of its output targets within the input targets provided by
$R_a$ towards channel $c_i$.

\begin{equation*}
\begin{split}
    T^{-}_{TCDG_{R_a}}(c_i) & \subseteq T^{+}_{TCDG_{R_b}}(c_i) \\
    \{A, B\} & \subseteq \{A, B\}
\end{split}
\end{equation*}

Cases will arise where facing with non target conforming local routing
function requires to extend $R_a$ (or $R_b$) by adding
routing choices.

\begin{definition}
    A \emph{routing function extension} to a routing function is
    valid if no deadlock may arise from the addition of new routing
    choices. The extended version of $R_a$ is denoted as $R_a'$ (a.k.a.
    extended $R_a$).

    In the context of a given channel $c$, the extension of $R_a$
    for channel $c$ results in $R_a'$ such that new routing choices
    are added for packets whose head flit is at the head of channel
    $c$. Thus,

    \begin{equation*}
        \bigcup_{d \in P} R_a(c, d) \subseteq \bigcup_{d \in P} R_a'(c, d)
    \end{equation*}

    This also applies for local routing functions at channel $c$:

    \begin{equation*}
        \bigcup_{d \in P} Local_{R_a\{c\}} \subseteq \bigcup_{d \in P} Local_{R_a'\{c\}}
    \end{equation*}

\label{def:rf-extension}
\end{definition}

\begin{definition}
    A \emph{routing function reduction} $R_a''$ of a routing function
    $R_a$ is valid if the removal of routing choices such that $\forall
    c \in C$, $Local_{R_a''\{c\}}$ is target conforming with respect to
    $R_a$ according to Definition \ref{def:target-conforming-lrf}. $R_a$
    can be considered as the extended version of $R_a''$. Thus, it can
    be expressed as $R_a = (R_a'')'$. In consequence, $R_a$ yields the
    same properties stated in Definition \ref{def:rf-extension} with
    respect to $R_a''$.
\label{def:rf-reduction}
\end{definition}

In the context of a given channel $c$, the reduction of $R_a$ for
channel $c$ results in $R_a''$ (a.k.a. reduced $R_a$) such that
existing routing choices are removed for packets whose head flit is
at the head of channel $c$.

\begin{definition}
    \emph{Compatible local routing function}. A non target conforming
    local routing function $Local_{R_b\{c\}}$ with respect to $R_a$ is
    said to be compatible with respect to $R_a$ for a given channel $c$,
    if a target conforming local routing function can be found from an
    extended version of $R_b'$ at channel $c$ with respect to either
    $R_a$ or an extended version $R_a'$ at channel $c$.
\label{def:compatible-lrf}
\end{definition}

\section{Related Work}\label{sec:related-work}

Dynamic reconfiguration schemes can be classified according to multiple
aspects. Following the taxonomy presented in \cite{casado2001protocol},
two main aspects are taken into account: deadlock avoidance and
unroutable packet management.

The schemes presented in this section employ different techniques to
avoid reconfiguration-induced deadlocks. We have identified two main
strategies:

\begin{itemize}
\item DAC: Deadlock avoidance by preventing the creation of cyclic
dependencies among network resources during the reconfiguration process.
\item DSR: Deadlock avoidance by splitting network resources into two
sets and statically reconfiguring those sets in sequence.
\end{itemize}

There is a third strategy for deadlock avoidance which allows the
creation of dependency cycles among network resources during the
reconfiguration process, but ensures that these cyclic dependencies do
not prevent forward progress of the reconfiguration process. Thus, at
some point during the process, these cycles will disappear avoiding
permanent deadlocks. Some routing algorithms benefit from this
approach\cite{schwiebert2001deadlock}, but, due to its complexity,
existing deadlock-free dynamic reconfiguration proposals do not use this
approach.

During the reconfiguration process, packets may reach channel queues
not allowed by the current routing function, these packets cannot be
routed because the routing function did not expect those packets to be
there (a.k.a. unroutable packets). If not handled, they can exhaust
network resources leading to performance degradation and possibly
deadlock due to permanent dependencies which cannot be removed. Three
strategies dealing with unroutable packets are considered in
\cite{casado2001protocol}:

\begin{itemize}
\item UD: Discard unroutable packets.
\item UB: Buffer unroutable packets as long as necessary to allow new
routing options to be available for these packets.
\item UR: Allow temporary routing options for unroutable packets. These
routing options may not be present by the end of the reconfiguration
process.
\end{itemize}

Table \ref{tab:related-work} presents the aforementioned dynamic
reconfiguration schemes according to the previous aspects.
PPR\cite{casado2001protocol} is a topology-agnostic reconfiguration
scheme which does not halt user traffic and it does not need additional
resources (such as virtual channels) to work. Instead, PPR relies on
the Up*/Down*\cite{schroeder1991autonet} routing algorithm to configure
escape paths. NetRec\cite{avresky2001dynamic} focuses on permanent node
faults. In order to restore connectivity, it builds a tree that spans
all immediate neighbors of the faulty node, whilst, packets routed
towards the faulty node are discarded.

\begin{table}[!t]
\renewcommand{\arraystretch}{1.3}
\caption{Dynamic Reconfiguration Schemes}
\label{tab:related-work}
\centering
\begin{tabular}{|c||c|c|c|}
\hline
& UD & UB & UR \\
\hline
\hline
DAC & PPR\cite{casado2001protocol}, NetRec\cite{avresky2001dynamic}, OSR\cite{lysne2008efficient},
       & uDIRECT\cite{parikh2015resource} & \emph{UPR}\\
    & OSRLite\cite{balboni2013optimizing},
        Lysne's\cite{lysne2000fast}, \emph{UPR} &
      Lysne's\cite{lysne2000fast} & \\
\hline
DSR & Double Scheme\cite{pinkston2003deadlock} & & \\
\hline
\end{tabular}
\end{table}

OSR\cite{lysne2008efficient} (and its modified version
OSRLite\cite{balboni2013optimizing}) is aimed to handle fault scenarios
as well as any other situation where paths are required to change. It
does not need additional resources neither relies on any particular
routing algorithm. OSR by design requires drainage of packets using
the old routing function at every channel in the network in order to
guarantee forward progress of the reconfiguration process. In the event
of faults, it performs drainage of unroutable packets by discarding
them.

Lysne's\cite{lysne2000fast} reconfiguration scheme objective is
to reduce the part of the network where a restructuring of the
Up*/Down* routing is necessary by only performing modifications on the
\emph{skyline} (tree's roots and intermediate nodes) of a multiple
root tree. If combined with PPR\cite{casado2001protocol} to trigger
the \emph{skyline} reconfiguration of the multiple roots tree it may
require discarding unroutable packets. However, if a synchronized
reconfiguration of the \emph{skyline} is performed, unroutable packets
remain stored. 

Another approach is uDIRECT\cite{parikh2015resource}, which handles
failures within unidirectional links. It makes use of a tree structure
to place deadlock-free escape routes on each fault scenario. In order
to prevent reconfiguration-induced deadlocks, unroutable packets are
ejected to the network interface of the router at which they are stored.

On the other hand, Double Scheme\cite{pinkston2003deadlock} uses two
sets of virtual channels in the network to perform spatial separation of
escape resources. It also does not impede the transmission of user
packets during the reconfiguration process, but it allows packet
discarding in case of disconnectivity from the packet perspective, i.e.
according to the resource where the packet is stored.

All the previous reconfiguration schemes provide an escape channel set
for packets in a predefined fashion to avoid reconfiguration-induced
deadlocks. In other words, the escape channel set for packets is
computed before the reconfiguration process takes place, and it remains
fixed during the whole process. Once packets enter escape channels, they
must remain within the escape channel set towards their destinations,
otherwise, deadlocks may arise. In order to achieve this, packets using
escape channels which may later use channels not in the escape channel
set must be drained.

In OSR, for instance, the escape channel set is given by the new routing
function. For this reason, all channels with packets using the old
routing function must be drained progressively before packets using
the new routing function can make use of them. In Double Scheme (and
its improved versions), the escape channel set is computed before the
reconfiguration process starts, by using a separate virtual channel
which must be drained from old packets before it can be used by new
packets.

Finally, our approach uses an hybrid strategy when dealing with
unroutable packets. It does allow temporary routing options in order
to enable dependencies to route unroutable packets towards their
destinations by exploiting existing compatibilities between the old
and new routing functions. Hence, the escape channel set is initially
given by the new routing function. However, as the reconfiguration
process progresses, new routing choices are added (if necessary) to the
new routing function to provide packets routed under the old routing
function, a valid escape path within the new routing function (they
would become unroutable packets otherwise).

As a consequence, in UPR, the escape channel set is dynamically updated
during the reconfiguration process, exploiting compatibility between
routing functions progressively in a channel basis, and reducing the
amount of channels which require drainage of old packets. Lower channel
drainage results in lower interference between packets routed under
different routing functions. Thus, packets routed under the new routing
function would not need to wait for drainage of packets routed under
the old routing function. This results in lower packet latency penalty
during the reconfiguration process.


\section{Upstream Progressive Reconfiguration}\label{sec:upr}

Briefly, UPR working principle is based on progressively performing
local reconfiguration at routers following the $TCDG$ associated
with the final routing function (or a modified version) in reverse
topological order, from sink channels towards source channels. Let
us consider the initial routing function $R_S$ before starting the
reconfiguration process $RP$. When the $RP$ is started, $R_P$ (the
prevailing routing function) matches $R_S$. Nevertheless, as the process
makes forward progress, $R_P$ evolves according to the modifications
made locally at routers for each channel. On the other hand, we denote
as $R_F$ to the final routing function (and its associated $TCDG_{R_F}
= G(C, D_{R_F})$ according to Definition \ref{def:tcdg}), which must be
configured once $RP$ finishes.

Graph traversal of $TCDG_{R_F}$ from sink channels towards source
channels allows the addition of new routing choices to $R_P$ (i.e.
dependencies in $TCDG_{R_P}$), enabling routes provided by $R_F$ to
packets being routed under the previous routing function. Following this
order, we can reduce the required amount of channels which need to be
drained of packets due to the lack of alternative routing choices. More
precisely, along the graph traversal, channels $Local_{R_F\{c\}}$ are
checked against Definition \ref{def:target-conforming-lrf} with respect
to $R_P$ for each visited channel $c$. Then, channel $c$ updates its
$Local_{R_P\{c\}}$ accordingly.

Additionally, in order to allow a more efficient RP progress,
UPR exploits compatibility between $R_F$ and $R_P$ locally at
routers according to Definitions \ref{def:rf-reduction} and
\ref{def:compatible-lrf}. Thus, UPR may perform changes on $R_P$ and/or
$R_F$ to exploit this compatibility. The resulting routing function
after these modifications to $R_F$ is an intermediate routing function
called $R_I$. Initially, $R_I = R_F$.

For the sake of clarity, updates to be performed on $R_P$ in successive
steps during the $RP$ result in a new routing function to be applied,
which we denote as $R_N$. Thus, after the corresponding operations
are performed, $R_N$ will become $R_P$ for later modifications (if
any). Each succeeding routing function has an associated $TCDG$.

As depicted, during the $RP$, each channel performs a series of steps
$step_i = <cond_i, oper_i>$ before $RP$ finishes. Thus, $cond_i$ is a
predicate on $c$. Similarly, $oper_i$ makes changes to the local routing
function affecting only to output choices related with channel $c$ at
the corresponding router $r$ (i.e. $c$ is an input channel of $r$, $c
\in Input(r)$).

\subsection{Step condition}

Assuming that a partial ordering is followed on channels, as we will
describe later, each channel $c$ checks $cond_i$ associated with
$step_i$. As formally defined below, channel $c$ must ensure that its
associated local routing function derived from $R_I$ (initially, $R_I
= R_F$) is a target conforming local routing function according to
Definition \ref{def:target-conforming-lrf} with respect to $R_P$ for
channel $c$.

\begin{condition}
    \emph{Target conforming $Local_{R_I\{c\}}$}.  A given channel $c$
    is allowed to perform $oper_i$ associated with $step_i$ if and only
    if $Local_{R_I\{c\}}$ is a target conforming local routing function
    with respect to $R_P$ for channel $c$.
\label{cond:target-conforming-lrf}
\end{condition}

\subsection{Step operation}

When upgrading channel $c$, routing choices from $R_I$ are added whilst
those from $R_P$ which do not conform (i.e. not present in $R_I$) are
removed for channel $c$. This operation is performed locally at each
router $r$ in a channel basis where $c \in Input(r)$. Hence, only
routing choices provided by $Local_{R_P\{c\}}$ are removed. Similarly,
routing choices corresponding to $Local_{R_I\{c\}}$ associated with
channel $c$ are added.

\begin{figure}[!t]
\centering
\includegraphics[width=\columnwidth]{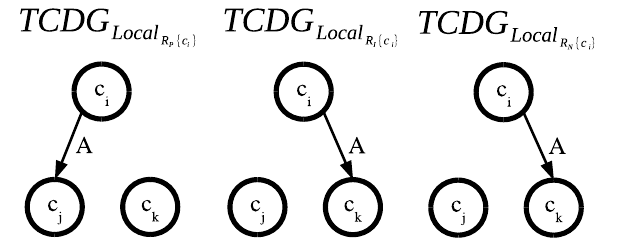}
\caption{Resulting $Local_{R_N\{c_i\}}$ for channel $c_i$ as a result
of applying the upgrade operation. Notice that $Local_{R_I\{c_i\}}$ is
target conforming with respect to $R_P$ for channel $c_i$, and so is
$Local_{R_N\{c_i\}}$.}
\label{fig:step-operation}
\end{figure}

Figure \ref{fig:step-operation} shows the resulting
$Local_{R_N\{c_i\}}$ for channel $c_i$ as a result of applying the
upgrade operation. Notice that $Local_{R_I\{c_i\}}$ is target conforming
with respect to $R_P$ for channel $c_i$, and so is $Local_{R_N\{c_i\}}$.

The step operation can be represented also using $R_P$ and $R_I$
from a global perspective. Further details are provided in Appendix
\ref{sec:step-operation-global} within the supplementary material.

\subsection{Reconfiguration order}\label{sec:reconfiguration-order}

\begin{figure}[!t]
\centering
\includegraphics[width=\columnwidth]{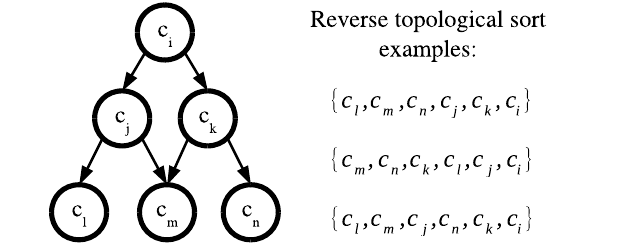}
\caption{Reverse topological sort example. For the same graph, multiple
sorts are shown.}
\label{fig:reverse-topological-order}
\end{figure}

The order in which routers carry out the corresponding $step_i$ is
the reverse topological order of $TCDG_{R_I}$ for each channel. Thus,
at the beginning, sink channels (i.e. $\forall c \in C :
D^+(c) = \emptyset$) check if $cond_i$ is satisfied. Figure
\ref{fig:reverse-topological-order} shows an example.

Sink channels $c_{sink}$ having at least one incoming dependency in
$TCDG_{R_I}$, i.e. $\forall c \in C : D^{+}_{TCDG_{R_I}}(c) = \emptyset
\land D^{-}_{TCDG_{R_I}}(c) \neq \emptyset$ ($\{c_l, c_m, c_n\}$ in
Figure \ref{fig:reverse-topological-order}), may start checking $cond_i$
as they have no successors in $TCDG_{R_I}$. If no topology changes
have been performed, these channels belong to $C_D$ (see Definition
\ref{def:in}).

That being said, sink channels do not satisfy Condition
\ref{cond:target-conforming-lrf} because the lack of output targets in
the $TCDG_{R_I}$. As a consequence, $Local_{R_I\{c_{sink}\}}$ will not
conform with respect to $R_P$ for sink channel $c_{sink}$:

\begin{equation*}
\begin{split}
    T^{-}_{TCDG_{R_P}}(c_{sink}) & \nsubseteq T^{+}_{TCDG_{R_I}}(c_{sink})\\
    T^{-}_{TCDG_{R_P}}(c_{sink}) & \nsubseteq \emptyset
\end{split}
\end{equation*}

Sink channels represent an exception which must be accounted
for. Therefore, Condition \ref{cond:target-conforming-lrf} is extended
as follows:

\begin{condition}
    \emph{Target conforming $Local_{R_I\{c\}}$ with sink exception}.  A
    given channel $c$ is allowed to perform $oper_i$ associated with
    $step_i$ if and only if $Local_{R_I\{c\}}$ is a target conforming
    local routing function with respect to $R_P$ for channel $c$ or
    $D^+(c) = \emptyset \land D^-(c) \neq \emptyset$.
\label{cond:target-conforming-lrf-sink}
\end{condition}

Source channels $c_{source}$ having at least one outgoing dependency
in $TCDG_{R_P}$, i.e. $\forall c \in C : D^{-}_{TCDG_{R_P}}(c) =
\emptyset \land D^{+}{TCDG_{R_P}}(c) \neq \emptyset$ ($c_i$ in Figure
\ref{fig:reverse-topological-order}) will always satisfy Condition
\ref{cond:target-conforming-lrf} regardless of their dependencies in
$TCDG_{R_I}$ because they do not have any incoming targets. Thus,
$Local_{R_I\{c_{source}\}}$ will always conform with respect to $R_P$
for source channel $c_{source}$:

\begin{equation*}
\begin{split}
    T^{-}_{TCDG_{R_P}}(c_{source}) & \subseteq T^{+}_{TCDG_{R_I}}(c_{source})\\
    \emptyset & \subseteq T^{+}_{TCDG_{R_I}}(c_{source})
\end{split}
\end{equation*}

Besides, orphan channels $c_{orphan}$ in $\{c: D^{-}_{TCDG_{R_P}}(c) =
\emptyset$ and $D^{+}_{TCDG_{R_P}}(c) = \emptyset, \forall c \in C\}$
(i.e. with no incoming/outgoing dependencies in $TCDG_{R_P}$) always
satisfy Condition \ref{cond:target-conforming-lrf} regardless of their
dependencies in $TCDG_{R_I}$. Orphan channels follow the same reasoning
as with source channels.

\subsection{Selective halting}\label{sec:selective-halting}

Upon channel $c$ condition evaluation, $c$ may not satisfy Condition
\ref{cond:target-conforming-lrf-sink}. Then, there exists at least one
incoming target to $c$ routed under $R_P$ for which is not provided any
alternative route by $R_I$.

\begin{equation*}
    T^{-}_{TCDG_{R_P}}(c) \nsubseteq T^{+}_{TCDG_{R_I}}(c)
\end{equation*}

In order to comply with Condition \ref{cond:target-conforming-lrf-sink},
it must be ensured that no incoming targets from $R_P$ for which $R_I$
does not provide any route at $c$ are fed into $c$ (a.k.a. offending
targets). This requires $c$ predecessors in $TCDG_{R_P}$ to remove
incoming dependencies (i.e. drain packets) towards $c$ bringing the
offending targets up to the source channels which inject packets towards
these targets going through $c$.

As a consequence, some pairs of processing nodes will be unable
to communicate temporarily due to the lack of connectivity among
them. On the other hand, injection will be halted for as short as
possible. This approach is referred to as \emph{selective halting}
dynamic reconfiguration, which corresponds to Case 7 in Lysne's
methodology\ \cite{lysne2005methodology}.

\begin{figure*}[!t]
\centering
\includegraphics[width=\textwidth]{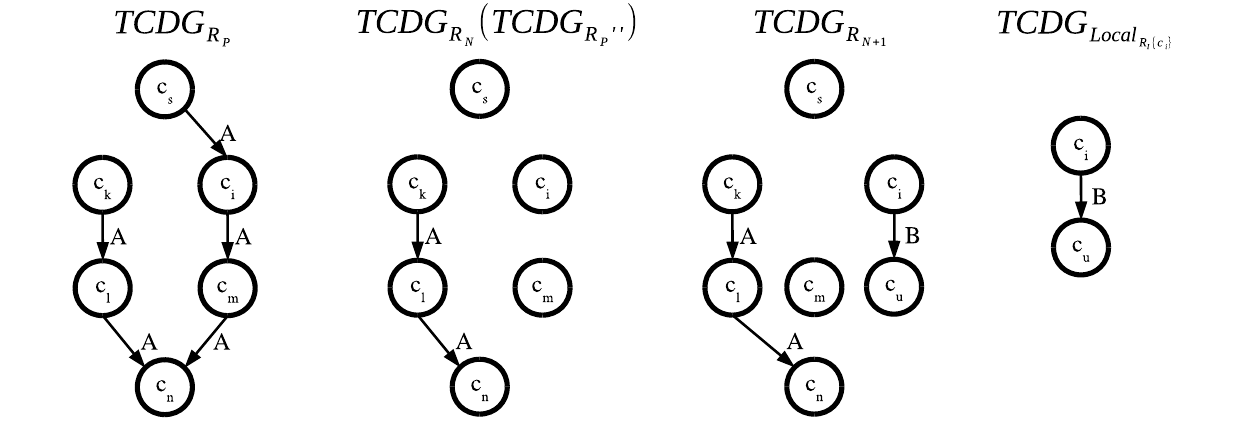}
\caption{\emph{Selective halting} example of offending target $A$
injected through sink channel $c_s$. $Local_{R_I\{c_i\}}$ is not target
conforming with respect to $R_P$ due to offending target $A$ coming from
$c_s$. Hence, removal of dependency $(c_s, c_i, A)$ (and unnecessary
dependencies underneath) is performed, resulting in $R_P''$ which makes
$Local_{R_I\{c_i\}}$ target conforming with respect to $R_P''$. Channel
$c_i$ completes its upgrade operation at $R_{N+1}$}
\label{fig:selective-halting}
\end{figure*}

Figure \ref{fig:selective-halting} shows an example of a \emph{selective
halting} scenario. Channel $c_i$ satisfies the RP order (see Section
\ref{sec:reconfiguration-order}) but it does not satisfy Condition
\ref{cond:target-conforming-lrf-sink}:

\begin{equation*}
\begin{split}
    T^{-}_{TCDG_{R_P}}(c_i) & \nsubseteq T^{+}_{TCDG_{R_I}}(c_i)\\
    \{A\} & \nsubseteq \{B\}
\end{split}
\end{equation*}

We can address this issue by halting injection of target $A$ at channel
$c_s$. Effectively removing dependency $(c_s, c_i, A)$ as well as
unnecessary dependencies underneath, reducing $R_P$, which becomes
$R_P''$. After this modification, $Local_{R_{I}\{c_i\}}$ is target
conforming with respect to $R_N$ (equal to $R_P''$). Thus, channel $c_i$
will satisfy Condition \ref{cond:target-conforming-lrf-sink} when $R_N$
becomes $R_P$ in the next step.

\subsection{Exploiting $TCDG$ conformability}\label{sec:exploiting-conformability}

Exploiting conformability is, in fact, one of the simplest cases of
interoperability between routing functions which helps to reduce
\emph{selective halting} in some cases. UPR takes advantage of the
existing conformability between $TCDG_{R_P}$ and $TCDG_{R_I}$ when
either channels are checking their step condition (i.e. upgrading) or
processing requests from successor channels in $TCDG_{R_P}$ to remove
dependencies.

\subsubsection{Exploiting conformability through $R_P''$ to remove dependencies}\label{sec:exploiting-conformability-rp-remove-deps}

Exploiting conformability through $R_P''$ consists in taking advantage
of a target conforming routing local function defined over $R_P''$
(reduced version of $R_P$, see Definition \ref{def:rf-reduction}) with
respect to $R_P$ to enable dependency removal from $TCDG_{R_P}$ if
needed. This is the case for routing functions offering multiple routing
choices towards the same target.

\begin{figure*}[!t]
\centering
\includegraphics[width=\textwidth]{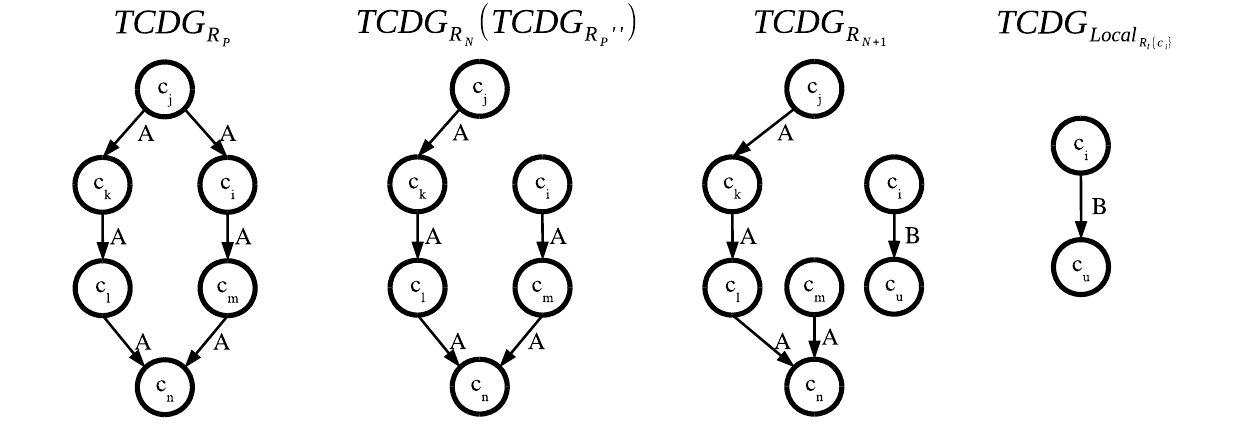}
\caption{Dependency $(c_j, c_i, A)$ removal exploiting conformability
due to alternative route provided by dependency $(c_j, c_k, A)$ at
$c_j$ for target $A$. Hence, $Local_{R_I\{c_i\}}$ is target
conforming with respect to $R_P''$ which will be used in the next step
(denoted as $R_N$). Finally, $c_i$ will be able to complete its upgrade
operation resulting in the graph denoted as $TCDG_{R_{N+1}}$.}
\label{fig:dep-removal-conformability-scenario}
\end{figure*}

Let us consider the scenario shown in Figure
\ref{fig:dep-removal-conformability-scenario}. Channel $c_i$ satisfies
the RP order (see Section \ref{sec:reconfiguration-order}) but
Condition \ref{cond:target-conforming-lrf-sink} is not met because
$Local_{R_{I}\{c_i\}}$ is not target conforming with respect to
$R_P$. Thus, $R_N$ would result in unroutable packets arising at channel
$c_i$ coming from $c_j$ towards target $A$.

Then, $c_i$ requests channel $c_j$ to remove dependency $(c_j,
c_i, A)$. Notice that the desired routing function $R_N$ (equal to
$R_P''$) to be applied next, yields $Local_{R_{N}\{c_j\}}$ to be
target conforming with respect to $R_P$. This is due to dependency
$(c_j, c_k, A)$ which provides an alternative route for target $A$ at
$c_j$. Then, $c_j$ can safely remove $(c_j, c_i, A)$ while keeping
routing connectivity. Hence, $c_j$ applies $Local_{R_{N}\{c_j\}}$ and it
acknowledges the removal to $c_i$.

Channel $c_i$ in turn, will apply its step operation once the
removal is acknowledged because it will satisfy Condition
\ref{cond:target-conforming-lrf-sink}. The resulting graph is shown as
$TCDG_{R_{N+1}}$.

\subsubsection{Exploiting conformability through $R_I''$ to upgrade channels}\label{sec:exploiting-conformability-ri-upgrade}

UPR may perform $R_I$ reduction enabling the RP to make progress
faster. To achieve this, a given channel $c_i$ may remove output choices
provided by $Local_{R_{I\{c_i\}}}$ towards a particular target $t$
if there exist alternative routes for which downstream channels in
$TCDG_{R_I}$ already completed their upgrade operation. Hence, removing
precedence restrictions towards non upgraded channels in $TCDG_{R_I}$.

\begin{figure}[!t]
\centering
\includegraphics[width=\columnwidth]{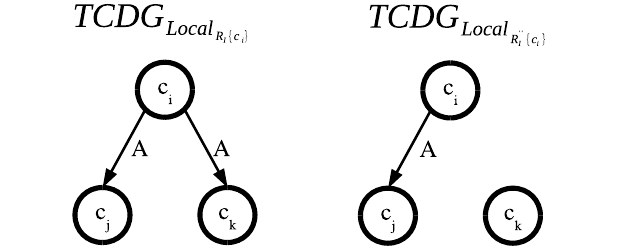}
\caption{Upon $c_j$ upgrades, following the reconfiguration order,
channel $c_i$ removes dependency $(c_i, c_k, A)$ to release from partial
order $c_k \prec c_i$. Thus, $c_i$ does not have to wait for $c_k$ to
complete its upgrade operation.}
\label{fig:upgrade-conformability-scenario}
\end{figure}

Figure \ref{fig:upgrade-conformability-scenario} shows a simple scenario
where conformability through $R_I''$ can be used. Channel $c_i$ cannot
check its upgrade condition until both channels $\{c_j, c_k\}$ complete
their upgrade operation due to the partial order established $\{c_j \prec
c_i, c_k \prec c_i\}$.

However, upon $c_j$ upgrade operation is completed, $c_i$ finds a target
conforming local routing function $Local_{R_I''\{c_i\}}$ with respect to
$R_I$ by removing dependency $(c_i, c_k, A)$ from $R_I$. Thus, $c_i$ may
start checking Condition \ref{cond:target-conforming-lrf-sink} because
now $c_i$ is only preceded by $c_j$.

\subsubsection{Restoring $R_F$ from $R_I''$}\label{sec:restoring-rf-from-reduced-ri}

Modifications performed on $R_I$ (initially, $R_I =
R_F$) due to conformability exploitation (see Section
\ref{sec:exploiting-conformability-ri-upgrade}) make $R_I$ to diverge
from $R_F$. Nevertheless, the reduced version $R_I''$ always remains
connected and it satisfies:

\begin{equation*}
    TCDG_{R_I''} \subseteq TCDG_{R_I} \subseteq TCDG_{R_F}
\end{equation*}

As explained in Section \ref{sec:exploiting-conformability-ri-upgrade},
routing choices removed from the initial $R_I$, were removed by channels
for which their direct successors in $TCDG_{R_I}$ were not upgraded to
speed up the RP. Hence, upon each direct successor completes its upgrade
operation, these routing choices can be restored. Therefore, they can
be added back to $TCDG_{R_I}$ safely without creating dependency cycles
assuming $R_F$ is deadlock-free.

The dependency restoration process starts from sink channels towards
source channels during the RP as soon as each channel completes its
upgrade operation following the reconfiguration order.

\subsection{Exploiting $TCDG$ compatibility}\label{sec:exploiting-compatibility}

Compatibility between two routing functions, namely $R_P$ and $R_I$
(initially $R_P = R_S$ and $R_I = R_F$), can be addressed from their
associated $TCDG$ perspective in a channel basis. The objective is to
reduce \emph{selective halting} at source channels.

As in Section \ref{sec:exploiting-conformability}, UPR takes advantage
of the existing compatibility between $TCDG_{R_P}$ and $TCDG_{R_I}$ when
channels are either checking their step condition or processing requests
from successor channels in $TCDG_{R_P}$ to remove dependencies.

\subsubsection{Exploiting compatibility through $R_P'$ to remove dependencies}\label{sec:exploiting-compatibility-rp-remove-deps}

\begin{figure*}[!t]
\centering
\includegraphics[width=\textwidth]{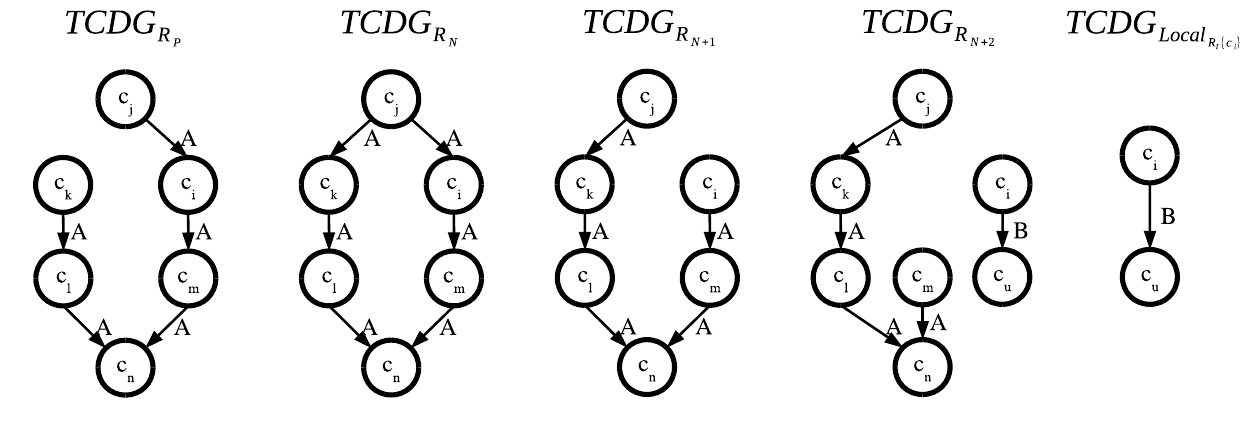}
\caption{Dependency removal compatibility scenario. Channel $c_i$
requests removal of dependency $(c_j, c_i, A)$. $c_j$ adds dependency
$(c_j, c_k, A)$ to $TCDG_{R_P}$ to fulfil the removal request.}
\label{fig:dep-removal-compatibility-scenario}
\end{figure*}

Let us consider the case for channel $c_i$ which satisfies
the reconfiguration order but it does not comply with
Condition \ref{cond:target-conforming-lrf-sink} due to the
offending target $A$. This scenario is shown in Figure
\ref{fig:dep-removal-compatibility-scenario}.

Channel $c_i$ requests $c_j$ to remove dependency $(c_j, c_i, A)$
bringing target $A$ into $c_i$ in $TCDG_{R_P}$. Upon processing the
removal request at $c_j$, it acknowledges that there exists an adjacent
channel in the topology $c_k$ which provides an alternative route
for target $A$ in $TCDG_{R_P}$. Besides, channel $c_k$ is not $c_j$
predecessor in $TCDG_{R_P}$. Otherwise, adding dependencies from $c_j$
to $c_k$ would create cycles in $TCDG_{R_P}$.

Then, by extending $R_P$ adding dependency $(c_j, c_k, A)$ (i.e.
becoming $R_P'$ in the next step ($R_{N}$)), $c_j$ will be able to
remove dependency $(c_j, c_i, A)$ by exploiting conformability. Thus,
performing $R_P'$ reduction $(R_P')''$, denoted as $R_{N+1}$ in Figure
\ref{fig:dep-removal-compatibility-scenario}. $Local_{R_{P}''\{c_j\}}$
(i.e. $R_P$ without dependency $(c_j, c_i, A)$) is a compatible local
routing function with respect to the initial $R_P$ according to
Definition \ref{def:compatible-lrf} because we found an extension $R_P'$
through the addition of dependency $(c_j, c_k, A)$ which is target
conforming with respect to $R_P$.

Besides, $Local_{R_{N+1}\{c_j\}}$ is target conforming with respect to
$R_N$. This modification, in turn, makes $Local_{R_I\{c_i\}}$ target
conforming with respect to $R_{N+1}$. Thus, $c_i$ will satisfy Condition
\ref{cond:target-conforming-lrf-sink} as soon as $R_{N+1}$ becomes
the prevailing routing function $R_P$. The resulting routing function
$R_{N+2}$ is shown after $c_i$ performs its upgrade operation.

\subsubsection{Exploiting compatibility through $R_I'$ to upgrade channels}\label{sec:exploiting-compatibility-ri-upgrade}

\begin{figure*}[!t]
\centering
\includegraphics[width=\textwidth]{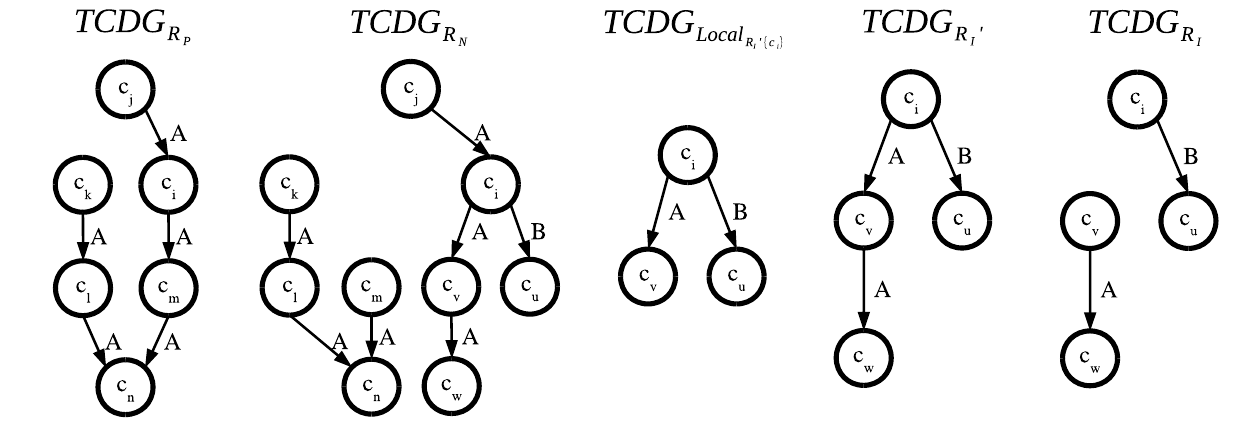}
\caption{Upgrade compatibility scenario. Channel $c_i$ extends $R_I$
adding dependency $(c_i, c_v, A)$ to provide alternative route for
target $A$ coming from $c_j$ in $R_P$. Hence, $Local_{R_I\{c_i\}}$ is
compatible with respect to $R_P$.}
\label{fig:upgrade-compatibility-scenario}
\end{figure*}

This case can be explained using the same scenario proposed
to exploit compatibility to remove dependencies, Figure
\ref{fig:upgrade-compatibility-scenario} illustrates this case. As in
Section \ref{sec:exploiting-compatibility-rp-remove-deps}, channel
$c_i$ satisfies the reconfiguration order but it does not comply
with Condition \ref{cond:target-conforming-lrf-sink} due to the
offending target $A$. Nevertheless, there exists an adjacent channel
in the topology $c_v$ providing an alternative route for target $A$
in $TCDG_{R_I}$. Additionally, $c_v$ is not $c_i$ predecessor in
$TCDG_{R_I}$. Otherwise, adding dependencies from $c_i$ to $c_v$ would
create cycles in $TCDG_{R_I}$.

According to Definition \ref{def:compatible-lrf}, $Local_{R_I\{c_i\}}$
is a compatible local routing function with respect to $R_P$ for
channel $c_i$ because it can extend $R_I$ (i.e. $R_I'$) to provide an
alternative route for target $A$. Thus, $Local_{R_N\{c_i\}}$ becomes
target conforming with respect to $R_P$.

Additionally, extending $R_I$ by adding dependency $(c_i, c_v, A)$ to
$TCDG_{R_I}$ also acts on $c_i$ upgrade order. In other words, now
$c_v \prec c_i$, as a result $c_i$ must wait for $c_v$ to upgrade
(unless $c_v$ was already upgraded). Deadlock may arise if exploiting
compatibility through $R_I$ is used in combination of conformability
exploitation through $R_I''$ when restoring missing dependencies
removed by the later. However, this can be avoided by taking removed
dependencies into account when extending $R_I$. Refer to Appendix
\ref{sec:deadlock-exploiting-conformability-compatibility} in the
supplementary material for more details.

\subsubsection{Restoring $R_F$ from $R_I'$}\label{sec:from-extended-ri-to-rf}

Exploiting compatibility through $R_I'$ to upgrade channels
is only performed by channels which do not satisfy Condition
\ref{cond:target-conforming-lrf-sink}. Therefore, there exists at least
one incoming target for which $Local_{R_{I}\{c\}}$ does not provide any
output choice. In other words, incoming targets exist at channels for
which $R_I$ did not expect to be there.

This means that eventually, channels bringing these incoming targets
using $R_P$ will complete their upgrade operation. Hence, old
dependencies towards upgraded channels downstream will be removed. Upon
these dependencies removal, channels which extended their local routing
function under $R_I$ (i.e. $Local_{R_{I}'\{c\}}$) will be able to remove
the previously added dependencies both in $R_P$ and $R_I$. Residual
dependencies on channels from older routing choices not provided by
$R_F$ are also known as \emph{ghost dependencies}.

In the worst case, \emph{ghost dependencies} will be removed at
source channels upon their upgrade operation is completed. Hence,
\emph{ghost dependencies} removal will propagate downwards $TCDG_{R_P}$
allowing upgraded channels to remove unnecessary dependencies both
in $TCDG_{R_P}$ and $TCDG_{R_I}$. The interested reader may refer
to Appendix \ref{sec:from-extended-ri-to-rf-example} in the
supplementary material for an example.

\subsection{Deadlock-free RP sufficient conditions}\label{sec:deadlock-free-conditions}

According to Lysne's methodology\cite{lysne2005methodology} a
deadlock-free reconfiguration process can be devised if the following
conditions are upheld:

\begin{enumerate}
\item The prevailing routing function is connected and deadlock-free at
all times.
\item During an \emph{Adding Phase}, no routing choice may be added to
any switch that closes a cycle of dependencies on escape resources.
\item During a \emph{Removing Phase}, no routing choice may be removed
from any switch before it is known that there will be no packet needing
this routing choice; either to proceed or to enter escape resources.
\item All potential \emph{ghost dependencies} are removed from the
network before a transition from a \emph{Removing Phase} to an
\emph{Adding Phase}.
\end{enumerate}

We assume both the initial routing function $R_S$ and final routing
function $R_F$ to be connected and deadlock-free before the RP starts
($1^{st}$ condition). During an \emph{Adding Phase} no routing choice
is added such that it creates a cycle on escape resources. This is
guaranteed by following the reconfiguration order explained in Section
\ref{sec:reconfiguration-order}.

Routing choices added by a given channel are added towards successor
channels which already completed their upgrade operation, therefore due
to the reconfiguration order being the reverse topological order, paths
towards \emph{sink} channels are always available through $R_I \subseteq
R_F$. Besides, routing choices added by exploiting compatibility are
added towards non predecessor channels in either $R_P$ (dependency
removal request) or $R_I$ (to satisfy the step condition). Hence, no
cycles can be formed during an \emph{Adding Phase} ($2^{nd}$ condition).

Channels may remove routing choices when applying their step
operation. In order to do that, they must fulfil the step condition,
which guarantees that incoming packets will always have a routing choice
alternative provided by $R_I$. Routing choices removed by exploiting
conformability/compatibility cannot remove dependencies such that
unroutable packets arise due to the computation of a $Local_{R_N\{c\}}$
which is target conforming with respect to $R_P$. Hence, during a
\emph{Removing Phase} it is guaranteed that packets will always have a
route available towards escape resources ($3^{rd}$ condition). In the
worst case scenario, \emph{selective halting} would be applied, draining
packets which would become unroutable otherwise before removing the
corresponding dependencies.

\emph{Ghost dependencies} are guaranteed to be either completely
removed or provisioned by exploiting compatibility through $R_I'$ on
channels which already completed their upgrade operation ($4^{th}$
condition). Thus, keeping escape resources free from dependency cycles
(\emph{ghost dependencies} removed) and unroutable packets (\emph{ghost
dependencies} provisioned in $R_I$).

\section{Evaluation}\label{sec:evaluation}

In this section, we evaluate the proposed reconfiguration scheme. The
goal of this evaluation is to assess the suitability of UPR and to show
that it can benefit from compatibility exploitation between two routing
algorithms to reduce \emph{selective halting} at sources while also
reducing the need for packet drainage at channels.

In the first part, we describe the methodology, topology and routing
algorithms used for the evaluation. Afterwards, we explain the metrics
used to present the results and analysis.

\subsection{Evaluation methodology}\label{sec:evaluation:methodology}

For our proposal evaluation we developed a sophisticated simulation
tool which allowed us to represent the reconfiguration process as a
concurrent computation of graph operations by multiple processes. Each
process comprises operations made by a single unidirectional channel
(channel for short) within the directed graph representing the network
topology.

Channel dependency information is available globally for all
channels. This information is stored using graph representations
$TCDG_{R_S}$, $TCDG_{R_P}$, $TCDG_{R_I}$ and $TCDG_{R_F}$ associated
to each routing function (see Section \ref{sec:upr} for more
details). These shared data structures yield a critical section and
they are accessed by channels to read and/or update their dependencies
information atomically. This guarantees dependency information to be
updated at all times so that decisions made by channels are always based
on the current state of the system.

Besides, each channel $c$ is in charge of the addition/removal of
outgoing dependencies devised from its local perspective. Therefore, $c$
will only perform modifications to $TCDG_{R_P}$, $TCDG_{R_I}$ involving
dependencies in $TCDG_{Local_{R_P\{c\}}}$, $TCDG_{Local_{R_I\{c\}}}$. In
other words, dependencies within sets $D^+_{TCDG_{R_P}}(c)$ and
$D^+_{TCDG_{R_I}}(c)$. This is consistent with Definition \ref{def:rf}
(a.k.a. distributed routing).

\subsubsection{Topology}

The network is built from a $5\times5$ mesh topology using bidirectional
physical links between router nodes. Each bidirectional link is
considered as two independent unidirectional channels. For simplicity,
we assume a single virtual channel (i.e. no virtual channels).

\subsubsection{Routing algorithms}

We consider four different minimal routing algorithms:
non-adaptive algorithms \emph{xy} and \emph{yx}, and the partially
adaptive algorithms \emph{odd-even}\cite{chiu2000odd} and
\emph{negative-first}\cite{glass1992turn}. Routing algorithms \emph{xy}
and \emph{yx} lack of any degree of adaptiveness, providing a single
route per source-destination pair. On the other hand, \emph{odd-even}
(\emph{oe}) and \emph{negative-first} (\emph{nf}) may provide multiple
routes for each source-destination pair and a reasonable degree of
adaptiveness.

Each combination of two distinct routing algorithms results in a
different scenario for the reconfiguration process to change from one
routing algorithm to the other.

\subsubsection{Evaluation metrics}

The first metric evaluated is the amount of channels which
require drainage of packets. We consider that a channel is
required to be drained of packets if it cannot satisfy Condition
\ref{cond:target-conforming-lrf-sink} due to some incoming
target/s. Thus, it has to request incoming dependencies removal to
upstream channels for one or more targets going through this channel
following the $TCDG_{R_P}$.

This metric considers a channel as \emph{requiring} drainage as
soon as a single target brought by the old routing function is
not provided any output choice by the new routing function. This
results in a conservative metric because it does not consider any
particular distribution of packets in the network. In a realistic
scenario, depending on the assignment of flits or packets to buffers,
channel drainage may not be necessary regardless of the output choices
provided by the new routing function. For example, neighbor traffic
will not distribute packets towards destinations comprising multiple
hops. Hence, regardless of the provided paths by the routing function
between source-destination pairs, some of them will never be used by
packets towards some destinations.

Another useful metric is the amount of halted flows (i.e.
source-destination pairs) for which \emph{selective halting} has been
applied during the RP for some period at source channels. As with
the previous metric, this considers a worst case scenario because
depending on the traffic pattern, some source-destination pairs may
never establish communication. In a real system, this would depend on
the job scheduling and task mapping techniques. Notice that injection
\emph{selective halting} is applied at some source-destination pairs if
and only if no available routes exist between those processing nodes.

\subsection{Results and analysis}\label{sec:evaluation:results-analysis}

\begin{figure*}[!t]
\centering
    \subfloat[]{
    \includegraphics[width=\thirdtextwidth]{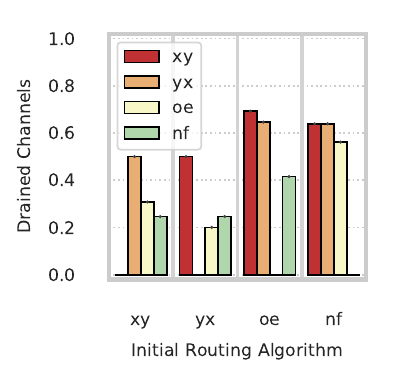}
    \label{fig:results:drained-channels:baseline}
    }
\hfil
    \subfloat[]{
    \includegraphics[width=\thirdtextwidth]{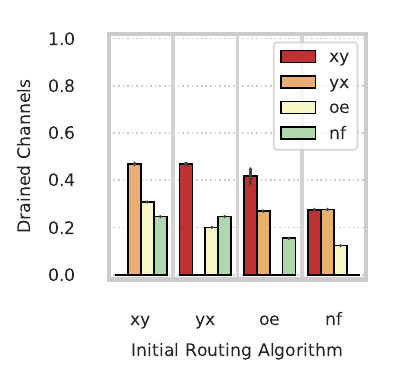}
    \label{fig:results:drained-channels:cp-ct-dcp-dct}
    }
\hfil
    \subfloat[]{
    \includegraphics[width=\thirdtextwidth]{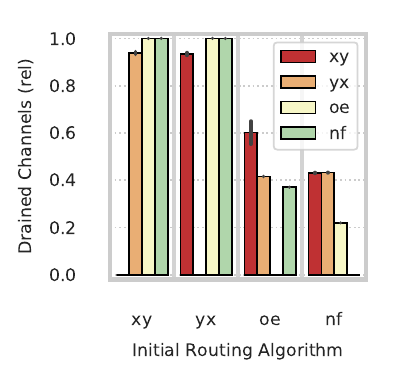}
    \label{fig:results:drained-channels-rel}
    }
\caption{Drained channels ratio: (a) Using only \emph{selective halting}
of traffic at source channels; (b) Exploiting conformability and
compatibility; (c) Exploiting conformability/compatibility with respect
to using \emph{selective halting} only.}
\label{fig:results:drained-channels}
\end{figure*}

The Upstream Progressive Reconfiguration algorithm has been evaluated
triggering the RP for each combination of routing algorithms. Figure
\ref{fig:results:drained-channels} shows the ratio of channels which
required drainage of at least one incoming target with respect to the
total amount of channels available in a $5\times5$ mesh network. Each
bar represents the final routing algorithm while the initial routing
algorithm is in the horizontal axis.

\begin{figure*}[!t]
\centering
    \subfloat[]{
    \includegraphics[width=\thirdtextwidth]{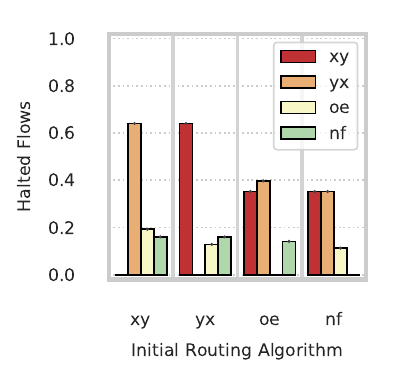}
    \label{fig:results:halted-flows:baseline}
    }
\hfil
    \subfloat[]{
    \includegraphics[width=\thirdtextwidth]{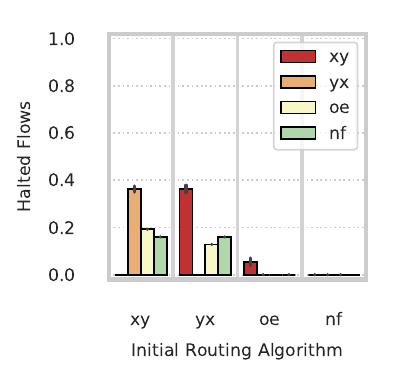}
    \label{fig:results:halted-flows:cp-ct-dcp-dct}
    }
\hfil
    \subfloat[]{
    \includegraphics[width=\thirdtextwidth]{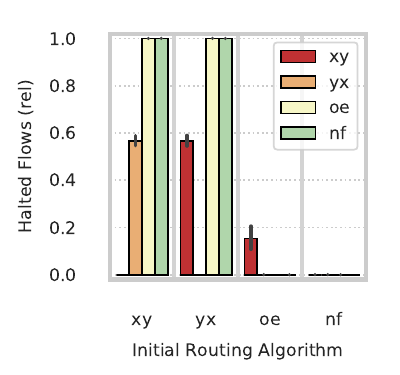}
    \label{fig:results:halted-flows-rel}
    }
\caption{Halted flows ratio: (a) Using only \emph{selective halting}
of traffic at source channels; (b) Exploiting conformability and
compatibility; (c) Exploiting conformability/compatibility with respect
to using \emph{selective halting} only.}
\label{fig:results:halted-flows}
\end{figure*}

Figure \ref{fig:results:drained-channels:baseline} shows the channel
ratio requiring drainage if relying solely in \emph{selective
halting}. First, the figure indicates that not all channels are
required to be drained, this is due to channels yielding a target
conforming $Local_{R_I\{c\}}$ with respect to $R_P$ (sink channels
always satisfy this). Non-adaptive algorithms \emph{xy} and \emph{yx}
have a single route per source-destination pair. Hence, when used as
initial routing algorithms, lower channel drainage is observed due to
these algorithms restraining the amount of different routes towards
the same destination.

On the other hand, using \emph{xy} and \emph{yx} as final routing
algorithms increases the drained channels ratio for the same reason. Due
to the existence of a single path between each source-destination
pair, the ratio of shared dependencies with the initial routing
algorithm is lower, specially when combined with partially adaptive
routing algorithms \emph{oe} and \emph{nf} as the initial routing
algorithms. Adaptiveness degree provided by \emph{oe} and \emph{nf} when
used as initial routing algorithms reduce the amount of channels with a
target conforming $Local_{R_I\{c\}}$ due to a higher amount of targets
going through channels at $R_P$. Thus, more channels will be drained in
these cases (greater than $60\%$).

Exploiting conformability and compatibility results
regarding drained channels are shown in Figure
\ref{fig:results:drained-channels:cp-ct-dcp-dct}. Clearly, this shows
great improvements when \emph{oe} and \emph{nf} are used as initial
routing algorithms due to the exploitation of conformability to reduce
$R_P$ in order to remove unwanted dependencies. In addition, when
combined with \emph{xy} and \emph{yx} as final routing algorithms, the
amount of drained channels is lowered to less than $45\%$ (lower than
$30\%$ for \emph{nf} as initial routing algorithm) due to the great
chances of exploiting compatibility within $R_I$ to upgrade channels,
increasing the amount of different routes provided by \emph{xy},
\emph{yx}. Greater reduction is obtained when combined with \emph{oe}
and \emph{nf} as final routing algorithms (lower than $20\%$) due to a
greater amount of shared dependencies (i.e. portions of paths between
source-destination pairs) which increases the amount of channels with
target conforming $Local_{R_I\{c\}}$.

Channel drainage ratio when exploiting conformability/compatibility
relative to just \emph{selective halting} is shown in Figure
\ref{fig:results:drained-channels-rel}. It can be observed that greater
improvements (i.e. less channels requiring drainage) are obtained
combining partially adaptive algorithms \emph{oe}, \emph{nf}. These
routing algorithms are maximally extended such that no dependencies
may be added to their associated $TCDG$ such that they do not create a
cycle. This increases the amount of target conforming $Local_{R_I\{c\}}$
for channels due to a large amount of shared dependencies among both
routing functions.

No significant improvement is obtained when algorithms \emph{xy},
\emph{yx} are used as initial routing algorithms with respect to relying
only on \emph{selective halting}. This is due to a low amount of target
conforming $Local_{R_I\{c\}}$ found among channels. Besides, due to
$oe$ and $nf$ being maximally extended routing functions, compatibility
cannot be exploited to a great extent when these algorithms are used as
final routing algorithms. Reduction of these routing algorithms would be
desirable in order to leave margin for UPR to add dependencies such that
non target conforming $Local_{R_I\{c\}}$ become compatible.

Channel drainage against other reconfiguration schemes can be greatly
reduced. For example, OSR always requires all channels to be drained
from all packets under $R_P$ for the proposed combinations of initial
and final routing algorithms. This is due to the use of the two distinct
sets of escape resources provided by $R_P$ and $R_I$. The former must
be used by packets routed under $R_P$ and the later will be used for
new injected packets under $R_I$. In OSR, at any given moment in time,
channels may belong to only one of these predefined escape sets. This,
in turn, makes OSR unlikely to benefit from the dynamic nature of the
reconfiguration process and existing compatibilities between these two
escape channel sets.

Figure \ref{fig:results:halted-flows} shows halted flows during the
reconfiguration process. Figure \ref{fig:results:halted-flows:baseline}
shows halted flows resulting from applying only \emph{selective
halting}. Worst cases are combinations between \emph{xy} and \emph{yx}
requiring to halt injection of more than $60\%$ of flows due to the lack
of alternative routes between each source-destination pair. Hence, a
single channel requiring the removal for a particular incoming target
triggers the injection \emph{selective halting} of that target at
multiple source channels.

Regardless of the initial routing algorithm used, setting \emph{oe} and
\emph{nf} as the final routing algorithms, keeps the ratio of halted
flows under $20\%$. This is a direct consequence of the increased amount
of routing choices (and portions of paths) which are shared with the
initial routing algorithm due to a greater number of paths between each
source-destination pair. Thus, a low amount of removal requests is
received at upstream channels to remove offending targets towards
downstream channels.

When exploiting conformability and compatibility halted flows results
are shown in Figure \ref{fig:results:halted-flows:cp-ct-dcp-dct}. We
can observe a great reduction in halted flows required from
the combination between \emph{xy} and \emph{yx}, which now
is less than $40\%$ which represents a $60\%$ with respect to
applying only \emph{selective halting} according to Figure
\ref{fig:results:halted-flows-rel}. Besides, halted flows for
\emph{oe} and \emph{nf} used as the initial routing algorithms are
completely avoided in some cases (i.e. $0\%$) except for combination
\emph{oe}-\emph{xy} which has been brought from around $37\%$ to $8\%$.

Overall, halted flows ratio is reduced when exploiting
conformability/compatibility. However, halted flows reduction
is greater for combinations among routing algorithms offering
multiple routes between source-destination pairs. This increases the
probability that some paths (or portion of paths) followed by packets
towards its destination can be shared among both routing algorithms,
either directly or by finding a compatible $Local_{R_I\{c\}}$ at
channels depending on the extension capability yielded by the final
routing algorithm.

Existing reconfiguration schemes do not rely on selective injection
halting of flows. Instead, they rely on buffer occupancy backpressure,
which may prevent source channels to inject packets towards any
destination. This is of special importance because old packets
interference with new packets may block new packets forwarding at a
given channel. This interference is propagated backwards, resulting in
new packets being blocked at source channels.

Proposals such as OSRLite\cite{balboni2013optimizing} try to alleviate
this problem by allowing packets routed under the new routing function
to be routed (from a certain point in the network) using the old
routing function. However, under a congested scenario this dramatically
increases the amount of packets under the old routing function that have
to be drained from the escape channel set provided by the new routing
function. Thus, delaying RP forward progress significantly.

\section{Conclusion}\label{sec:conclusion}

In this paper, we have proposed a new process for deadlock-free dynamic
reconfiguration which is able to exploit existing conformability and
compatibility among two distinct routing functions. This process is
applicable to any routing algorithm independently of the underlying
topology. Besides, it can be applied to packet switching, virtual
cut-through and wormhole switching (the interested reader may refer
to \cite{duato2005theory} for a formal proof), and it does not
require additional resources. Our proposal guarantees the absence
of deadlocks during the reconfiguration process relying on Duato's
theory\cite{duato2005theory}.

The results show that it is not always necessary to drain all channels
from packets using previous routing choices but they can be rerouted
according to the new routing function as long as it provides alternative
route choices for those packets.

What is more interesting is that both the old routing function and
the new routing function can be modified in a partially ordered
stepped process such that they coexist. Coexistence among routing
functions reduces interference of the reconfiguration process with
the transmission of packets. This can lead to scenarios where
uninterrupted transmission of packets between processing nodes is
allowed while the reconfiguration process takes place.

Another important aspect regarding the potential benefits obtained
by UPR is that conformability can be exploited to a higher degree
when the initial routing algorithm provides a greater amount of paths
(i.e. routing choices) between source-destination pairs. On the other
hand, compatibility exploitation can be achieved when the final
routing algorithms provide a smaller amount of routing choices. This
puts on display the potential of this algorithm to be applied for
reconfiguration scenarios due to planned/unplanned channel deactivation
which may result in topology changes (e.g. energy saving policies,
component failures, etc.).

%
Finally, UPR addresses the challenges introduced when reconfiguring
networks. It reduces the amount of channels requiring drainage of
packets while also reducing packet injection halting at processing
nodes. Additionally, it avoids reconfiguration-induced deadlocks while
performing routing choice addition/removal on a channel-by-channel
basis.

\section{Future work}\label{sec:future-work}

Our findings suggest the potential use of UPR to address multiple issues
involving network reconfiguration. Besides, there are some aspects of
our proposal which can be improved to achieve better performance and/or
extended to provide support for different scenarios. A brief description
of these aspects is provided in this section.

%
First, better results are expected if additional resources are provided
for UPR to exploit conformability and compatibility with respect to
channel drainage and halted flows. This is similar to the working
principle of reconfiguration techniques such as the Fully Adaptive
Double Scheme\cite{pinkston2003deadlock}.

An interesting use case of UPR is the integration with
\emph{power-aware} techniques to improve energy efficiency. Results show
the potential of UPR to perform planned addition/removal of routing
choices (possibly due to on/off links) while reducing interference with
packet transmission. UPR always ensures escape path availability prior
to dependencies removal due to link shutdown (and all its associated
channels, including virtual channels). This provides \emph{power-aware}
techniques full flexibility.

On the contrary, \emph{power-aware} techniques which solely rely on
routing algorithms adaptiveness degree, can lead to an incomplete escape
channel set. As a consequence, unroutable packets may arise resulting in
deadlock configurations. Thus, diminishing \emph{power-aware} techniques
flexibility.

Besides, further analysis of the use of UPR in conjunction with topology
agnostic routing algorithms such as Tree-turn Routing\cite{zhou2012tree}
or Segment-based Routing\cite{mejia2006segment} in order to deal with
topology changes would be desirable. These issues will be addressed
further in future work.

Last, ongoing research is being developed to provide unplanned link
deactivation support (e.g. component failures). Unplanned link
deactivation may result in unroutable packets being blocked at channel
queues. Existing proposals handle this scenario discarding those
packets. Thus, preventing reconfiguration-induced deadlocks while
guaranteeing forward progress of the reconfiguration process.

Due to UPR compatibility exploitation, alternative escape resources
can be provided to blocked packets by extending either $R_P$ or
$R_I$. Hence, reducing the amount of packets which have to be
discarded. Nevertheless, this must be performed carefully, such that
these extensions do not prevent forward progress of the reconfiguration
process.

\ifCLASSOPTIONcompsoc
  \section*{Acknowledgments}
\else
  \section*{Acknowledgment}
\fi

This work has been jointly supported by the Spanish MINECO and European
Commission (FEDER funds) under the project RTI2018-098156-B-C52
(MINECO/FEDER), and by Junta de Comunidades de Castilla-La Mancha under
the project SBPLY/17/180501/000498. Juan-Jos\'{e} Crespo is funded by
the Spanish MECD under national grant program (FPU) FPU15/03627.

\clearpage

\appendices
\section{Global representation of step operation}\label{sec:step-operation-global}

From the global routing function point of view, the resulting $R_N$
applied at next step can be induced from its $TCDG_{R_N}$ representation
which can be formulated using graph operations as combination of
$TCDG_{R_P}$ and $TCDG_{R_I}$. First, we compute the induced subgraph on
$TCDG_{R_I}$ by $c_i$ direct successors (out-neighborhood $N^+(c_i)$)
including $c_i$, denoted as $TCDG_{R_I\{c_i\}}$.

\begin{equation*}
    TCDG_{R_I\{c_i\}} = TCDG_{R_I}[N_{TCDG_{R_I}}^{+}(c_i) \cup \{c_i\}]
\end{equation*}

Then, we remove $c_i$ outgoing dependencies from $TCDG_{R_P}$. The
resulting graph is denoted as $TCDG_{R_P\{-c_i\}}$.

\begin{equation*}
    TCDG_{R_P\{-c_i\}} = (C, \{(c_j, c_k, t) : (c_j, c_k, t) \in D_{R_P} \land c_j \neq c_i\})
\end{equation*}

\begin{figure*}[!t]
\centering
\includegraphics[width=\textwidth]{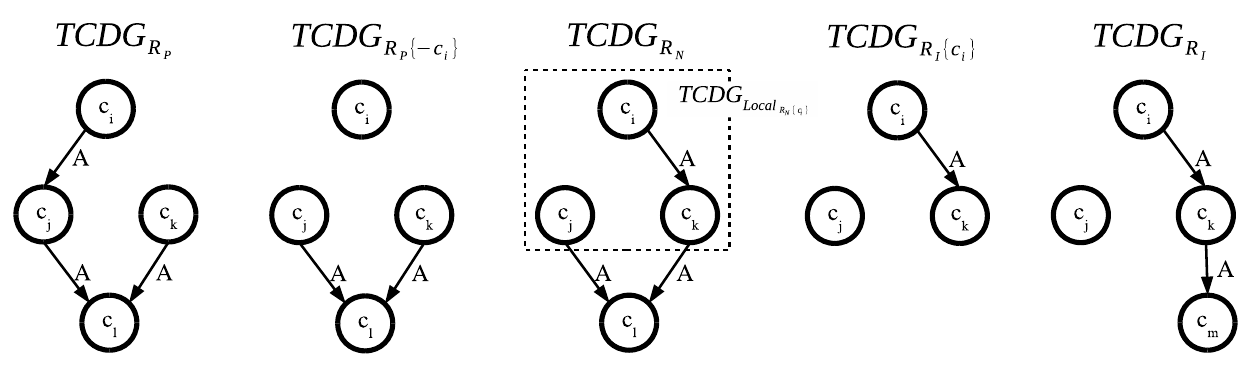}
\caption{Example of $TCDG_{R_N}$ computation. Only the framed portion
($Local_R(R_N, r)$) of the graph matches $TCDG_{R_I}$, while the
remaining dependencies are taken from $TCDG_{R_P}$.}
\label{fig:step-operation-global}
\end{figure*}

Finally, $TCDG_{R_N}$ is obtained adding graphs $TCDG_{R_P\{-c_i\}}$ and
$TCDG_{R_I\{c_i\}}$. Figure \ref{fig:step-operation-global} illustrates
an example of $TCDG_{R_N}$ computation.

\begin{equation*}
    TCDG_{R_N} = TCDG_{R_P\{-c_i\}} + TCDG_{R_I\{c_i\}}
\end{equation*}

\section{Deadlock avoidance when exploiting conformability through
$R_I''$ while exploiting compatibility through $R_I'$ to upgrade
channels}\label{sec:deadlock-exploiting-conformability-compatibility}

There is an important consideration that must be taken into
account when exploiting conformability through $R_I''$ (see
Section \ref{sec:exploiting-conformability-ri-upgrade}) and also
compatibility through $R_I'$ to upgrade channels, in order to prevent
reconfiguration-induced deadlocks.

\begin{figure*}[!t]
\centering
    \subfloat[Initial setup.]{
    \includegraphics[width=\columnwidth]{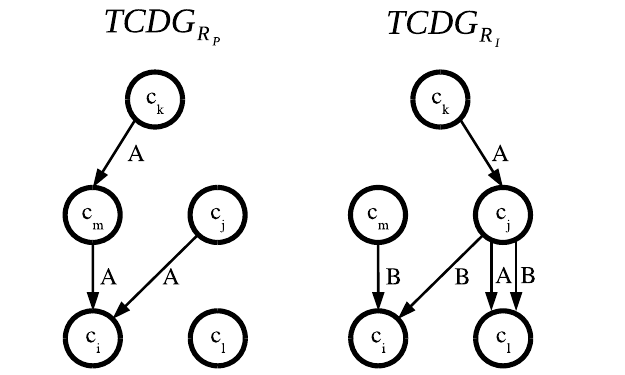}
    \label{fig:deadlock-upgrade-compatibility-conformability:setup}
    }
\hfil
    \subfloat[A cycle arises following the reconfiguration order $\{c_l,
    c_j, c_k, c_i\}$. Channel $c_j$ removes dependency $(c_j, c_i, B)$
    to comply with Condition \ref{cond:target-conforming-lrf-sink}
    in $TCDG_{R_{N+1}}$. After $c_i$ completes its upgrade operation,
    $c_j$ restores dependency $(c_j, c_i, B)$ resulting in
    $TCDG_{R_{N+4}}$ (the dependency cycle is shown in red).]{
    \includegraphics[width=\textwidth]{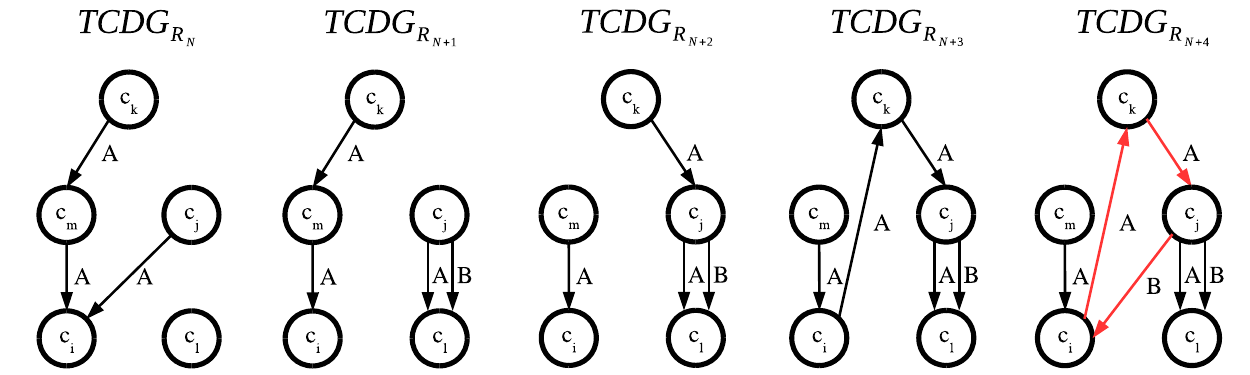}
    \label{fig:deadlock-upgrade-compatibility-conformability:trace}
    }
\caption{Deadlock example scenario when restoring $R_F$ from $R_I''$ due
to a previous removal by exploiting conformability through $R_I''$ while
exploiting compatibility through $R_I'$. Only a portion of the involved
$TCDGs$ is shown.}
\label{fig:deadlock-upgrade-compatibility-conformability}
\end{figure*}

Given a channel $c_i$ with a compatible $Local_{R_I\{c_i\}}$ which
extends $R_I$ by adding some dependency towards an adjacent channel
in the topology $c_k$. Channel $c_k$ may not be $c_i$ predecessor
in $TCDG_{R_I}$ but it could be in $TCDG_{R_F}$ because $R_I$ is a
reduced version of $R_F$. Then, when restoring $R_F$ from $R_I''$ (see
Section \ref{sec:restoring-rf-from-reduced-ri}) by adding missing
dependencies, cycles may arise. Let us see an example scenario in Figure
\ref{fig:deadlock-upgrade-compatibility-conformability} by tracing
a possible computation of a $RP$.

Initially, all channels $\{c_i, c_j, c_k, c_l, c_m\}$
in $TCDG_P$ are not upgraded. We also assume that these
channels are neither sources nor sinks (see Figure
\ref{fig:deadlock-upgrade-compatibility-conformability:setup}). Only
a portion of the $TCDGs$ is shown for simplicity. Hence, additional
dependencies to keep both $R_P$ and $R_I$ connected are assumed and
omitted for brevity.

The RP trace is shown in Figure
\ref{fig:deadlock-upgrade-compatibility-conformability:trace}. First,
channel $c_l$ completes its upgrade operation resulting in
$TCDG_{R_N}$. In the next step, $c_j$ realizes that it could satisfy
the reconfiguration order by exploiting conformability through
$R_I''$ .Thus, removing dependency $(c_j, c_i, B)$ from $TCDG_{R_I}$,
giving $TCDG_{R_{N+1}}$ as a result. Afterwards, following the
reconfiguration order, channel $c_k$ upgrades and the result is shown in
$TCDG_{R_{N+2}}$.

Later, channel $c_i$ checks Condition
\ref{cond:target-conforming-lrf-sink} which does not satisfies
due to target $A$ brought by $c_m$. It extends $R_I$ by adding
dependency $(c_i, c_k, A)$, completing its upgrade operation (see
$TCDG_{R_{N+3}}$). Due to $c_i$ upgrade completion, $c_j$ restores
dependency $(c_j, c_i, B)$ in $TCDG_{R_{N+4}}$. A cycle comprising
channels $\{c_j, c_i, c_k\}$ arises.

Two different approaches can be devised to prevent
reconfiguration-induced deadlocks in this scenario:

\begin{enumerate}
\item When \emph{exploiting compatibility through $R_I'$ to upgrade
    channels}: Prevent dependency addition towards $c_i$ predecessors in
    $TCDG_{R_I} + TCDG_{R_F}$. This assumes that missing dependencies
    from $TCDG_{R_F}$ can be added at any moment to $TCDG_{R_I}$.

\item When \emph{restoring $R_F$ from $R_I''$} (see Section
    \ref{sec:restoring-rf-from-reduced-ri}): Delay dependency
    restoration towards channel $c_i$ if cycles arise until removal of
    auxiliary dependencies introduced by $R_I'$.
\end{enumerate}

For simplicity, we have considered the first approach in UPR's
implementation.

\section{Restoring $R_F$ from $R_I'$ example}\label{sec:from-extended-ri-to-rf-example}

\begin{figure*}[!t]
\centering
    \subfloat[Reached state during the RP. Dashed arrows represent extra
    dependencies added to $R_I$ that do not exist in $R_F$. In this
    case, dependency $(c_i, c_m, A)$ was added due to incoming target
    $A$ into $c_i$.]{
    \includegraphics[width=\textwidth]{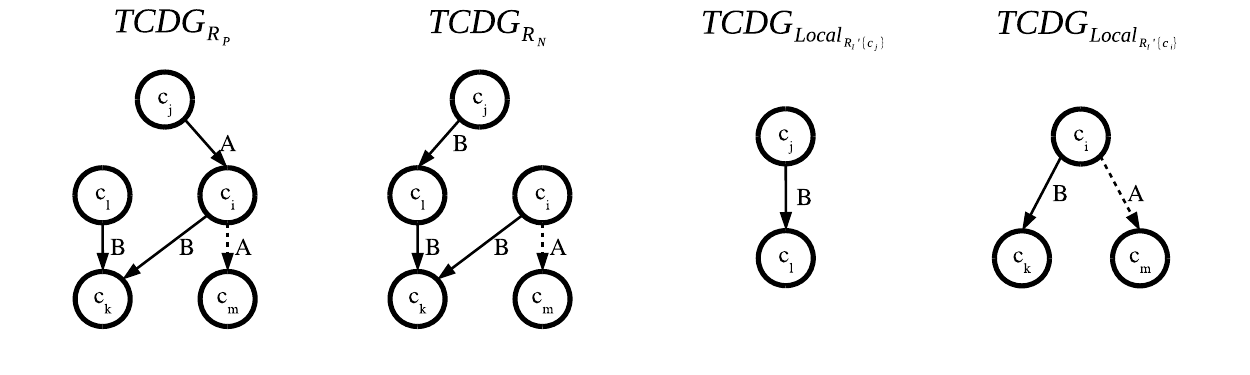}
    \label{fig:restoring-rf-from-extended-ri:setup}
    }
\hfil
    \subfloat[Channel $c_j$ completes its upgrade operation. In
    consenquence, it removes dependency $(c_j, c_i, A)$, allowing for
    the extra dependency (dashed arrow) to be removed from both $R_P$
    and $R_I$.]{
    \includegraphics[width=\textwidth]{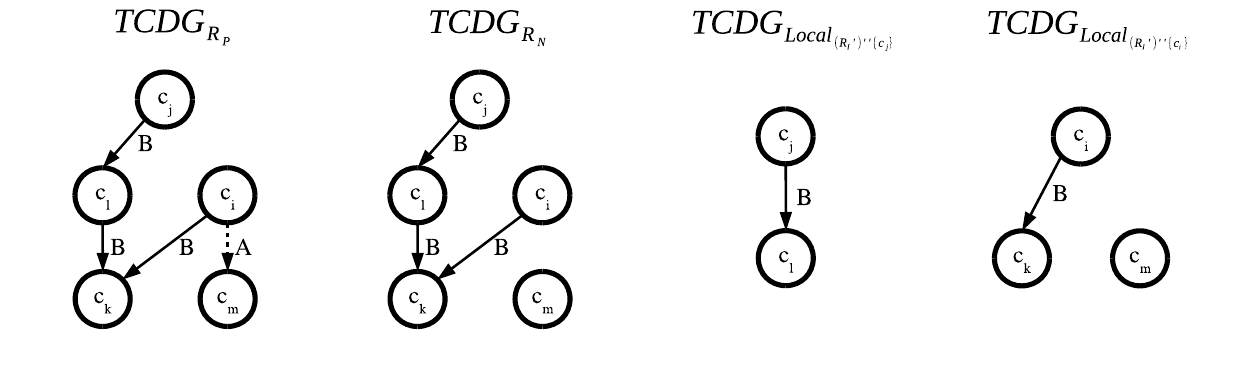}
    \label{fig:restoring-rf-from-extended-ri:solved}
    }
\caption{Restoring $R_F$ from $R_I'$ example.}
\label{fig:restoring-rf-from-extended-ri}
\end{figure*}

Figure \ref{fig:restoring-rf-from-extended-ri} illustrates an example
scenario where extra dependencies added during the RP to $TCDG_{R_I}$
are removed. Recall that these dependencies were introduced by
exploiting compatibility through $R_I'$ to upgrade channels (see
Section \ref{sec:exploiting-compatibility-ri-upgrade}). Channel
$c_i$ added dependency $(c_i, c_k, A)$ (dashed in Figure
\ref{fig:restoring-rf-from-extended-ri:setup}) to satisfy Condition
\ref{cond:target-conforming-lrf-sink} and it completed its upgrade
operation in a previous step. Thus, $c_i$ is applying an extended local
routing function $Local_{R_P}\{c_i\} = Local_{R_I'}\{c_i\}$ derived from
$R_I$. In the current state, upstream channel $c_j$ is performing its
upgrade operation. Therefore it satisfies the reconfiguration order and
Condition \ref{cond:target-conforming-lrf-sink} because either:

\begin{itemize}
\item $Local_{R_I}\{c_j\}$ is target conforming with respect to $R_P$.
\item Channel $c_j$ found a target conforming
$Local_{R_I'}\{c_j\}$. Therefore, $Local_{R_I}\{c_j\}$ is
compatible with respect to $R_P$ according to Definition
\ref{def:compatible-lrf}. \item Removal of dependencies bringing
target $A$ from upstream channels in $TCDG_{R_P}$.
\end{itemize}

Once $c_j$ completes its upgrade operation, removal of dependency
$(c_j, c_i, A)$ propagates towards $c_i$. Channel $c_i$ realizes
that there are no more incoming dependencies bringing target
$A$. Therefore, $c_i$ removes the previously added dependency $(c_i,
c_k, A)$ from $TCDG_{R_P}$ and $TCDG_{R_I}$ because dependency
$(c_i, c_k, A)$ does not belong to $TCDG_{R_F}$ (see Figure
\ref{fig:restoring-rf-from-extended-ri:solved}). Dependency removal will
keep propagating towards downstream channels in $TCDG_{R_P}$.

\clearpage
\clearpage



%

\bibliographystyle{IEEEtran}
\bibliography{IEEEabrv,bibliography.bib}

\end{document}